\documentclass[12pt]{iopart}

\usepackage{iopams}  

\usepackage{graphicx}

\usepackage{setstack}
\usepackage[ruled,linesnumbered,boxed]{algorithm2e}

\usepackage{natbib}
\usepackage{har2nat}
\setcitestyle{aysep={}}
\usepackage[backref, colorlinks=true, citecolor=blue, urlcolor=blue, linkcolor=blue]{hyperref}

\begin{document}

\title[Simulating Photometric Images of Moving Targets with Photon-mapping]{Simulating Photometric Images of Moving Targets with Photon-mapping}

\author{Junju Du, Shaoming Hu, Xu Chen, Hai Cao and Yuchen Jiang}

\address{Shandong Key Laboratory of Optical Astronomy and Solar-Terrestrial Environment, \\
	School of Space Science and Physics, Institute of Space Sciences,\\ 
	Shandong University,  Weihai, Shandong, 264209, China.}
\ead{husm@sdu.edu.cn}
\vspace{10pt}
\begin{indented}
\item[]\today
\end{indented}

\begin{abstract}
We present a novel, easy-to-use method based on the photon-mapping technique to simulate photometric images of moving targets. Realistic images can be created in two passes: photon tracing and image rendering. The nature of light sources, tracking mode of the telescope, point spread function (PSF), and specifications of the CCD are taken into account in the imaging process. Photometric images in a variety of observation scenarios can be generated flexibly. We compared the simulated images with the observed ones. The residuals between them are negligible, and the correlation coefficients between them are high, with a median of $0.9379_{-0.0201}^{+0.0125}$ for 1020 pairs of images, which means a high fidelity and similarity. The method is versatile and can be used to plan future photometry of moving targets, interpret existing observations, and provide test images for image processing algorithms. 
\end{abstract}

%
\vspace{2pc}
\noindent{\it Unified Astronomy Thesaurus concepts}: Astronomical simulations (1857); Astronomy data modeling (1859); Artificial satellites (68), Space debris (1542); Near-Earth objects (1092)

%
\submitto{Publications of the Astronomical Society of the Pacific}
%
%
%

\section{Introduction}\label{sec:introduction}
In the optical survey of moving targets, such as satellites, space debris, and near-Earth objects (NEOs), the photometric images have different characteristics depending on the different observation strategies. \citeasnoun{Kouprianov.2010} divided all images into four types according to the shape of the source images: 

(1) point-like field stars and point-like targets;

(2) point-like field stars and streak-like targets; 

(3) streak-like field stars and point-like targets;

\begin{figure}[ht!]
    \centering
    \includegraphics[width=1.0\columnwidth]{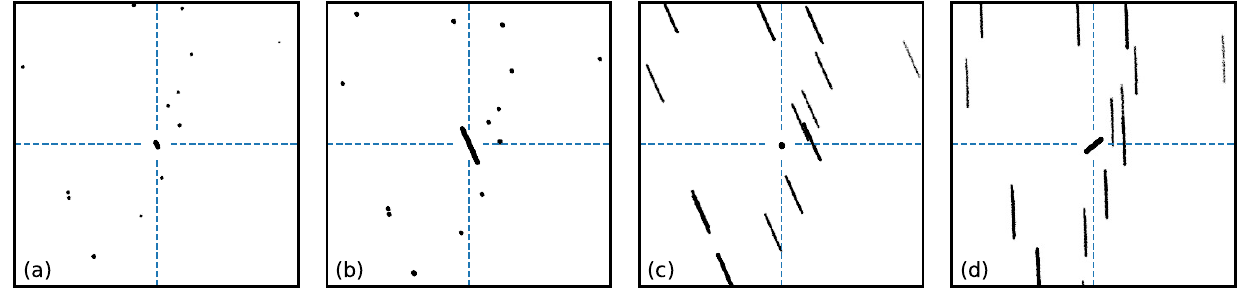}
    \caption{Schematic diagrams of four types of photometric images. Images are simulated with the presented method in different strategies. (a) sidereal tracking mode and 0.5 seconds exposure. (b) sidereal tracking mode and 4.0 seconds exposure. (c) target tracking mode and 4.0 seconds exposure. (d) parking tracking mode and 4.0 seconds exposure.  The moving target is indicated by the intersecting dashed lines.\label{fig:figure01}}
\end{figure}

\begin{figure}[ht!]
	\centering
	\includegraphics[width=0.6\columnwidth]{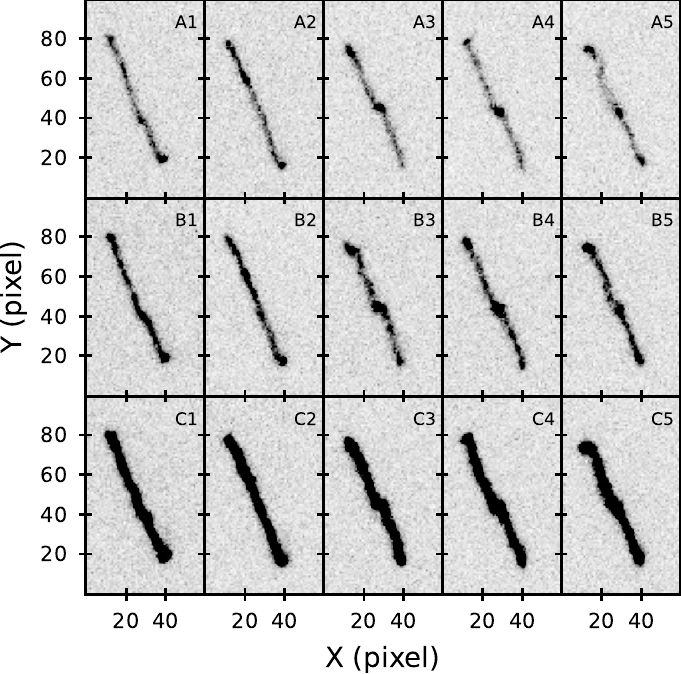}
	\caption{Screenshots of three streak-like field stars (A, B, C) in five consecutive frames (1, 2, 3, 4, 5). The exposure times are 2 seconds, and the interval between adjacent images is about 6 seconds. All images were obtained with the one-meter telescope at Weihai Observatory of Shandong University.\label{fig:figure02}}	
\end{figure}

(4) streak-like field stars and streak-like targets.\\
The four types of photometric images are illustrated in Figure~\ref{fig:figure01}. The first type of image has the obvious advantage that it does not require any specialized image processing techniques. However, its practical use is limited to moving targets with an apparent motion similar to stars. A short exposure time option is another way to obtain the first type of images, but only for very bright stars and targets. Streak-like images are difficult to avoid in the observation of moving targets. Three factors complicate the image processing of streak-like images. First, the signal-to-noise ratio (SNR) is lost, and the boundary is blurred since the stars or targets become trailed~\cite{Veres.2012}. Second, as shown in Figure~\ref{fig:figure02}, the shapes of streak-like images change unpredictably from frame to frame and cannot be expressed in an analytic form accurately. Third, streak-like images are more likely to overlap each other or cross image boundaries, which can be seen in Figure~\ref{fig:figure01} (c) and (d)

Streak-like images can be handled appropriately by choosing the most suitable tracking mode of the telescope, setting the optimal exposure time, and developing more efficient image processing technology. However, many problems deserve further study in each aspect. The tracking mode is limited by the telescope's tracking mechanism and should be appropriate to the research target. Generally, the target tracking mode is used for follow-up observation, while the sidereal tracking mode is used for searching for new targets. Some new tracking strategies have been proposed, but their actual effects need to be verified in various observation scenarios. Setting the optimal exposure time is not easy either. As the telescope direction changes, some sources enter the field of view, and some run out. The exposure time needs to be set dynamically according to the sources within the field of view. The position and flux extraction of streak-like images is the fundamental problem. Various extraction algorithms have been proposed, such as thresholding segmentation~\cite{Devyatkin.2010}, profile fitting~\cite{Veres.2012,Fraser.2016}, and new ones periodically appear. These methods need to be tested under an identical and complete test image set to provide a basis for their applicable conditions and development direction.

Realistic image simulation is an effective means to deal with streak-like images in all three aspects mentioned above. By constructing a high-fidelity imaging model, we can obtain a moving target's virtual image in any given scene in advance. These images may help us to make appropriate decisions on tracking mode and exposure time. Moreover, they can act as data samples for image processing because their parameters are known and adjusted flexibly. Image simulation is not a novel subject. \citeasnoun{Massey.2004} presented a method to simulate deep sky images with shapelets, including realistic galaxy morphologies and telescope characteristics. \citeasnoun{Bertin.2009} developed an astronomical image simulation software package called SkyMaker, which started as a testing tool for the SExtractor source extraction software and has since been used in various studies. Unfortunately, none of them deals with moving targets and streak-like images. \citeasnoun{Peterson.2015} presented a comprehensive methodology PhoSim, which uses a photon Monte Carlo approach to calculate the ab initio physics of the atmosphere and a telescope \& camera to simulate realistic astronomical images. However, Phosim is not flexible in generating images with given parameters. Some models to generate test images are proposed in the literature, e.g., 1D stretched Gaussian~\cite{Kouprianov.2008} or moving 2D Gaussian~\cite{Veres.2012}. However, these analytic models can not characterize the discrete distribution of photons and the distortion of streak-like images very well, especially when the light sources are faint and the telescope is jitter.

In this paper, we adopted the strategy of photon-mapping to synthesize the optical images of moving targets. Photon-mapping is an efficient and versatile global illumination technique for realistic image synthesis. It has been developed in computer graphics in the last few years~\cite{Jensen.2001}. In general, photon-mapping involves two passes: photon tracing and image rendering. In the first pass, photons are traced from the light sources into the scene. These photons, which carry flux information, are cached in a data structure called the photon map. In the second pass, an image is rendered by evaluating each pixel's value using the information stored in the photon map and the property of the detector~\cite{Dutre.2006}. This method simulates the propagation and transformation of photons from the perspective of probability theory. It is closer to the actual imaging process, which we will prove in subsequent sections. 

To adapt to the specific problems in moving targets observation, we made some adaptive changes to the usual photon-mapping technique in global illumination. The whole image simulation framework is described in Section~\ref{sec:methods}. The validity of the framework is verified by comparing the simulated images with the observed ones in Section~\ref{sec:results}. The framework is practical, flexible, and efficient, but some subtle effects need to be considered in the future. Conclusions and discussion are given in Section~\ref{sec:discussion}. 

\section{Formation of the photometric image}\label{sec:methods}
We will analyze how a moving target and field stars are imaged on the telescope's CCD in this section.  Relative to a ground station, a moving target travels among filed stars with an apparent angular velocity.  During a given exposure time, photons emitted by the moving target and filed stars pass through the atmosphere, the telescope's lens set, and then arrive at the image plane. The photons captured by CCD are converted into analog-to-digital units (ADU) and read out as fits or other format images. The final image is determined by the nature of sources, the telescope's tracking mode, atmospheric conditions, and CCD characteristics. We will describe the process of image formation from the perspective of photon-mapping below. 

\subsection{Observation geometry}
\begin{figure}[ht!]
    \centering
	\includegraphics[width=0.6\columnwidth]{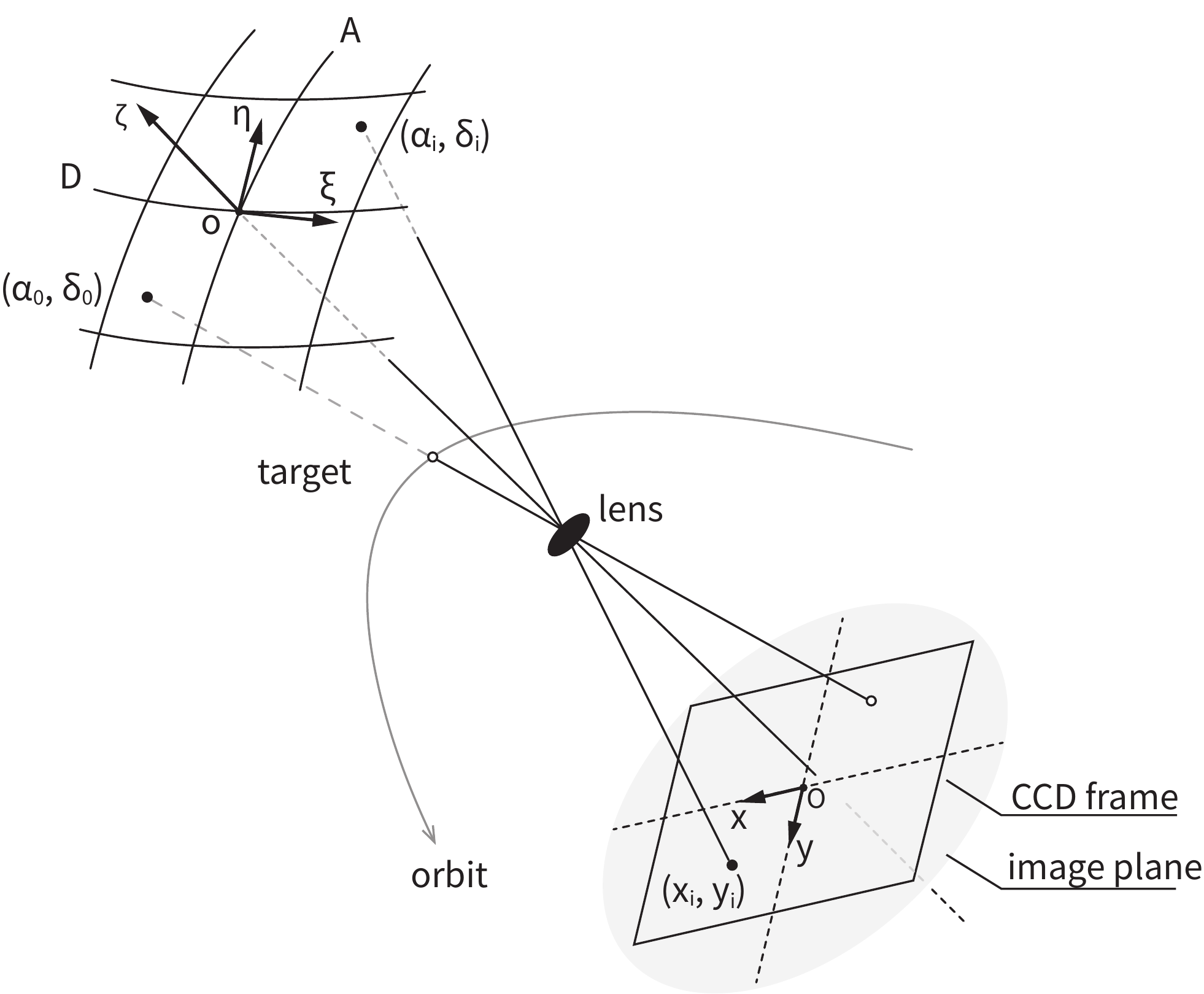}
	\caption{The geometry of photometric observation.}
	\label{fig:figure03}
\end{figure}

The observation geometry describes the spatial relationships between the moving target, the field stars, the ground station, and the telescope. The relationships change over time due to the Earth's rotation, the target's orbital motion, and the telescope's pointing adjustment. The analysis of observational geometry is essential because it determines the mapping relationship between the celestial coordinates ($\alpha, \delta$) and the image coordinates ($x,y$). These two coordinates are connected by a local transitional coordinate called standard coordinates ($\xi, \eta$). As shown in  Figure~\ref{fig:figure03}, the origin of standard coordinates is the intersection of the optical axis and the celestial sphere and the tangent of the plane $\xi \bi{o} \eta$ to the celestial sphere. The $\xi$ is directed towards the North pole, tangent to the local celestial meridian, and $\eta$ is perpendicular and directed eastward, parallel to the celestial equator. The transformation from ($\alpha, \delta$) to ($\xi, \eta$) is called gnomonic or central projection~\cite{Kovalevsky.2004}. There are 
\begin{eqnarray}
	\xi(\alpha, \delta, A, D)
	&=& \frac{\cos\delta\sin(\alpha-A)}{\sin\delta\sin D+\cos\delta\cos D\cos(\alpha-A)} \label{eq:xi_adad}\\
	\eta(\alpha, \delta, A, D)
	&=& \frac{\sin\delta\cos D-\cos\delta_i\sin D\cos(\alpha-A)}{\sin\delta\sin D+\cos\delta\cos D\cos(\alpha-A)},\quad \label{eq:eta_adad}
\end{eqnarray}
where ($A$, $D$) are right ascension and declination of bore-sight of the telescope, ($\alpha$, $\delta$) are those of field stars or the moving target in the vicinity of $(A, D)$. The plane of CCD (or image plane) does not coincide exactly with the focal plane due to the CCD installation's inaccuracy. The image coordinate is defined on the image plane, with the center of CCD as the origin and the directions parallel to the two edges of CCD as the ${\rm x}$ and ${\rm y}$ directions. The ($\xi, \eta$) and $(x,y)$ are linked by photographic plate models~\cite{Kovalevsky.2004}
\begin{eqnarray}
	\xi(x, y, \mathbf{P}) =
	     && +p_{1}x+p_{2}y+p_{3}+p_{5}x+p_{6}y\label{eq:xi_xyp}\\
	     && +p_{7}x^{2}+p_{8}xy+p_{9}x(x^2+y^2) \label{eq:xi_xyp_nolinear}\\
	\eta(x, y,\mathbf{P}) =
	     && -p_{2}x+p_{1}y+p_{4}+p_{6}x-p_{5}y\label{eq:eta_xyp}\\
	     && +p_7xy+p_{8}y^{2}+p_{9}y(x^2+y^2) \label{eq:eta_xyp_nolinear},
\end{eqnarray}
where $\mathbf{P} = \{p_{i}|i=1,2,\cdots,9\}$ are plate constants, which can be estimated in astrometry process. Parameters $p_1 \cdots p_4$ represent an orthogonal model allowing for a zero point, scale and rotation between ($x, y$) and ($\xi, \eta$). Adding parameters $p_5$ and $p_6$ gives a full linear model. Parameters $p_7$ and $p_8$ represent a tilt of the focal plane with respect to the ideal tangent plane, and $p_9$ is the third-order optical distortion term. In general, $\xi$ and $\eta$ are in radians while $x$ and $y$ are in pixels or microns. For a telescope with small field of view, the nonlinear terms represented by Equations~(\ref{eq:xi_xyp_nolinear}) and~(\ref{eq:eta_xyp_nolinear}) are very small and can be ignored. Equations~(\ref{eq:xi_xyp}) and~(\ref{eq:eta_xyp}) are one-value and have one-valued inverses
\begin{eqnarray}
	x &=& \xi^{-1}(\xi, \eta, \mathbf{P})\\
	y &=& \eta^{-1}(\xi, \eta, \mathbf{P}).
\end{eqnarray}

The ($\alpha, \delta$) in Equations~(\ref{eq:xi_adad}) and~(\ref{eq:eta_adad}) may be the coordinates of field stars or a moving target. For field stars, the ($\alpha, \delta$)  can be obtained by correcting their catalog positions at the epoch to the time of observation. For a moving target, the ($\alpha, \delta$) change rapidly and should be calculated with orbit prediction. At each instance of time $t$, the position vector of a moving target in Geocentric Celestial Reference System (GCRS), $\bi{X}_{\rm GCRS}(t)$, can be predicted by orbit elements and corresponding propagation models, such as the most commonly used two-line element sets (TLEs) and the SDP4/SGP4 models~\footnote{\url{https://www.space-track.org}}. For a ground station at geodetic coordinates (latitude $\phi$, longitude $\lambda$, height $h$), its corresponding coordinates in GCRS can be expressed as
\begin{eqnarray}\label{eq:Y_GCRS}
	\bi{Y}_{\rm GCRS}(t) = {W}(t){R}(t){M}(t)\bi{Y}_{\rm ITRF}(t,\phi,\lambda, h),
\end{eqnarray}
where $\bi{Y}_{\rm ITRF}(t,\phi,\lambda, h)$ is the station coordinate in International Terrestrial Reference System (ITRS). The components ${W}(t)$, ${R}(t)$ and ${M}(t)$ are the transformation matrices arising from the motion of the celestial pole owing to precession and nutation, the rotation of the Earth around the same pole, and polar motion, respectively~\cite{Kovalevsky.2004}. More information about the four terms in Equation~(\ref{eq:Y_GCRS}) is available in International Earth Rotation and Reference Systems Service (IERS)~\footnote{\url{https://www.iers.org}}. Therefore, the station-target vector can be written as 
\begin{eqnarray}
	{\brho}(t) &=\left[\rho_1, \rho_2, \rho_3\right]^{\rm T} = \bi{X}_{\rm GCRS}(t) - \bi{Y}_{\rm GCRS}(t).
\end{eqnarray}
The corresponding right ascension and declination of target respecting to the station can be obtained with
\begin{eqnarray}
	\alpha_0(t) &=& \arctan(\rho_2/\rho_1)\in [0, 2\pi]\\
	\delta_0(t) &=& \arcsin(\rho_3) \in [-\pi/2, \pi/2].
\end{eqnarray}
This paper will always use the subscript 0 to indicate the target, thus distinguishing it from the field stars. The above mathematical derivation gives a map from the given orbital elements and site coordinates to the station-target vector and ($\alpha_0,\delta_0$). In practical application, there are many off-the-shelf computer programs for the above calculation, such as the software routines from the IAU SOFA Collection~\footnote{\url{http://www.iausofa.org}} or Skyfield~\cite{Rhodes.2019}.

The $(A, D)$ in Equations~(\ref{eq:xi_adad}) and (\ref{eq:eta_adad}) changing over time refers to the tracking movement of telescope. Generally, the main purpose of tracking movement is to keep the light from the sources of interest always on the same spot on the image plane during the exposure~\cite{Eichhorn.1963}. The $A(t)$ and $D(t)$ can be arbitrary expressions as long as the telescope permits. We focus on three tracking modes from the perspective of practical application: (1) target tracking mode, (2) sidereal tracking mode, and (3) parking mode. Target and sidereal tracking modes mean the telescope is fixed on the moving target or a fixed point on the celestial sphere, respectively. Parking mode means the telescope's tracking mechanism shut down, and the telescope changes direction as the Earth rotates. However, during the tracking movement, a telescope is always influenced by wind, gravity, and thermal loads deforming~\cite{Andersen.2011}. There is an additional wobble as the telescope moves according to a given strategy. \citeasnoun{Andersen.2011} listed several servomechanism performance metrics to describe the telescope's tracking accuracy or instability. However, the wobble of a telescope is still hard to describe. In this paper, we focused on the uncorrectable tracking error. We modeled the tracking movement as a 2D Brownian motion in the right ascension and declination direction, similar to the method used by~\citeasnoun{Peterson.2015}. Therefore, there are
\begin{eqnarray}
	A(t) &=& \alpha_0(t) + \sigma_{\rm A}B_{\rm A}(t)+\epsilon_{\rm A} \label{eq:A_target}\\
	D(t) &=& \delta_0(t) + \sigma_{\rm D}B_{\rm D}(t)+\epsilon_{\rm D} \label{eq:D_target}
\end{eqnarray}
for target tracking mode,
\begin{eqnarray}
	A(t) &=& A_0 + \sigma_{\rm A}B_{\rm A}(t)+\epsilon_{\rm A}\\
	D(t) &=& D_0 + \sigma_{\rm D}B_{\rm D}(t)+\epsilon_{\rm D}
\end{eqnarray}
for sidereal tracking mode, and
\begin{eqnarray}
	A(t) &=& A_0+\sigma_{\rm A}B_{\rm A}(t)+\omega(t-t_0) + \epsilon_{\rm A} \label{eq:A_parking}\\
	D(t) &=& D_0+\sigma_{\rm D}B_{\rm D}(t)+\epsilon_{\rm D} \label{eq:D_parking}
\end{eqnarray}
for parking mode. Where ($\alpha_0(t), \delta_0(t)$) is the direction of the moving target, $B_{\rm A}(t)$ and $B_{\rm D}(t)$ are 1D standard Brownian motion in right ascension and declination direction, respectively, $\sigma_{\rm A}$ and $\sigma_{\rm D}$ are corresponding scaling parameter, ($A_0, D_0$) is the initial pointing direction of telescope, $\epsilon_{\rm A}$ and $\epsilon_{\rm D}$ are the initial pointing error, $\omega=15^{\prime\prime} {\rm s}^{-1}$ is the rotation speed of the Earth. The tracking path of telescope during an exposure is a sample curve of Brownian motion. A simulation of a continuous-time process is always based on a finite sample. Brownian motion is usually simulated on a discrete time grid $t_{0}\le t_{1}\le \cdots \le t_{n}$, and it is common practice for visualizations to interpolate the simulated values linearly~\cite{Schilling.2012}. Since a $d$-dimensional Brownian motion has independent 1D Brownian motion as components, we can use the Algorithm~(\ref{alg:algorithm01}) to generate the components $\Delta A(t) = \sigma_{\rm A}B_{\rm A}(t)+\epsilon_{\rm A}$ and $\Delta D(t) = \sigma_{\rm D}B_{\rm D}(t)+\epsilon_{\rm D}$ in Equations~(\ref{eq:A_target})-(\ref{eq:D_parking}) separately.

\begin{algorithm}\label{alg:algorithm01}
	\caption{The simulation of 2D Brownian motion.}
	\KwIn{A discrete time grid $t_{0}\le t_1\le\cdots \le t_k\le \cdots \le t_n \le t_0+t_{exp}$. The scale parameters $\sigma_{\rm A}$ and $\sigma_{\rm D}$ of 1D Brownian motion.}
	\KwOut{A sample curve of Brownian motion.}
	Initialize $\Delta A(t_0) \gets \epsilon_{\rm A}$ and $\Delta D(t_0) \gets \epsilon_{\rm D}$ \;
	\For {$k \gets 1$ \KwTo  $n$}{
	Generate ${\rmd}A \sim {\rm N}(0, (t_{k}-t_{k-1})\sigma_{\rm A}^2)$ \;
	Generate ${\rmd} D \sim {\rm N}(0, (t_{k}-t_{k-1})\sigma_{\rm D}^2)$ \;
	Set $\Delta A(t_k) \gets \Delta A(t_{k-1})+ {\rmd}A$ \;
	Set $\Delta D(t_k) \gets \Delta D(t_{k-1})+ {\rmd}D$ \;	
	}
	Return $\Delta A(t_k)$ and $\Delta D(t_k)$ for $k=1,\cdots, n$.
\end{algorithm}

\subsection{Photon tracking}
Photon tracing is the process of emitting discrete photons from the light sources and tracing them through the scene. This pass's primary goal is to populate the photon maps used in the rendering pass. The light sources emit photons, and the photons travel through space filled with a variety of media, some are absorbed, and some survive. We do not analyze how the photons interact with the atmosphere, telescope, and camera in detail, but only how the surviving photons are distributed on the image plane.

During an exposure from $t_0$ to $t_0+t_{\rm exp}$, the surviving photons strike into the image plane one by one. This process can be described with Poisson Point Processes (PPPs) in the time-space $[t_0, t_0+t_{\rm{exp}}]$. Suppose there are $N_{\rm s}$ field stars and one target that intersect the field of view during exposure, their celestial coordinates and magnitudes are $(\alpha_i, \delta_i)$ and $m_i$, respectively, with $i=0$ for target and $i=1,\cdots,N_{\rm s}$ for field stars. The parameters of field stars can be obtained from the star catalog, and the parameters of the target can be obtained from the orbital prediction and previous observations. The number of photons arriving at the image plane can be estimated by
\begin{equation}
    N_{{\rm p}\it{i}}  = t_{\rm exp} 10^{-0.4(m_i+K^{'}X(z)+ZP)}\label{eq:Np},
\end{equation}
where $K^{'}$ is the first-order extinction coefficients of atmosphere, $z$ is the zenith distance, $X(z)$ is the airmass, and $ZP$ is the zero point. The values of $K^{'}$ and $ZP$ are determined by the atmosphere, telescope, and filter, which are readily available for a particular set of observing devices. Such parameters as the telescope collecting area, the transmission of atmosphere and instrument, the fractional spectral bandwidth of the filter, and the source flux density are also used in literature to estimate the photon number~\cite{Peterson.2015}. However, it is not easy to obtain these parameters. For a device-specific simulation, Equation~(\ref{eq:Np}) is more convenient. A definite value of $N_{\rm{p}\it{i}}$ can be given by Equation~(\ref{eq:Np}) from $m_i$, or directly assigned as we want, which helps to generate an image with given flux information. When we need to consider the fluctuation of the photons number of the same source in different images, we can replace $N_{\rm{p}\it{i}}$ with a random number of events drawn from a Poisson distribution with mean $N_{\rm{p}\it{i}}$.

The photons hit the image plane one by one. The arrive time $t_{ij}$ of the $j$-th photon from $i$-th source can be considered as uniformly distributed in  $[t_0, t_0+t_{\rm{exp}}]$, with $i=0,\cdots, N_{\rm{s}}$ and $j=1,\cdots, N_{\rm{p}\it{i}}$. The coordinates of the photon falling into the image plane can be obtained with the following formulas
\begin{eqnarray}
	A_{ij} &=& A(t_{ij})\\
	D_{ij} &=& D(t_{ij})\\
	\alpha_{ij} &=& \alpha_{i}(t_{ij})\\
	\delta_{ij} &=& \delta_{i}(t_{ij})\\
	\xi_{ij} &=& \xi(\alpha_{ij}, \delta_{ij}, A_{ij}, D_{ij})\\
	\eta_{ij} &=& \eta(\alpha_{ij}, \delta_{ij}, A_{ij}, D_{ij})\\
	x_{ij} &=& \xi^{-1}(\xi_{ij}, \eta_{ij}, \mathbf{P}) \\
	y_{ij} &=& \eta^{-1}(\xi_{ij}, \eta_{ij}, \mathbf{P}) \\
	\tilde{x}_{ij} &=& x_{ij} + \epsilon_{\rm x} \\
	\tilde{y}_{ij} &=& y_{ij} + \epsilon_{\rm y},
\end{eqnarray}
where $(\epsilon_{\rm x}, \epsilon_{\rm y}) \sim PSF(x,y)$ are the deviations of the coordinates of the photon caused by PSF effect in ${\rm x}$ and ${\rm y}$ direction, respectively. \citeasnoun{Bertin.2009} assumed the PSF to be the convolution of five components: atmospheric blurring, telescope motion blurring, instrument diffraction and aberrations, optical diffusion effects, and intra-pixel response. However, we use the given PSF to generate $\epsilon_{\rm x}$ and $\epsilon_{\rm y}$. The PSF can be given as an analytical form, such as a 2D Gaussian function, or derived from observational fitting~\cite{Anderson.2000}.

As mentioned before, the information of each photon will be stored in a data structure called a photon map. In global illumination, the data structure of a kd-tree or a Voronoi diagram is widely used to improve access speed. But in this paper, the photons are all located in the image plane. A simple 2D array is enough to store the information of photons effectively. In a photon map, each row represents a photon, which contains the following fields: ($i$, $j$, $t_{ij}$,  $A_{ij}$, $D_{ij}$, $\alpha_{ij}$, $\delta_{ij}$, $\xi_{ij}$, $\eta_{ij}$, $x_{ij}$, $y_{ij}$, $\tilde{x}_{ij}$, $\tilde{y}_{ij}$). The meaning of each field has been introduced in the above. 

\subsection{Image rendering}
\begin{figure}[ht!]
    \centering
	\includegraphics[width=0.6\columnwidth]{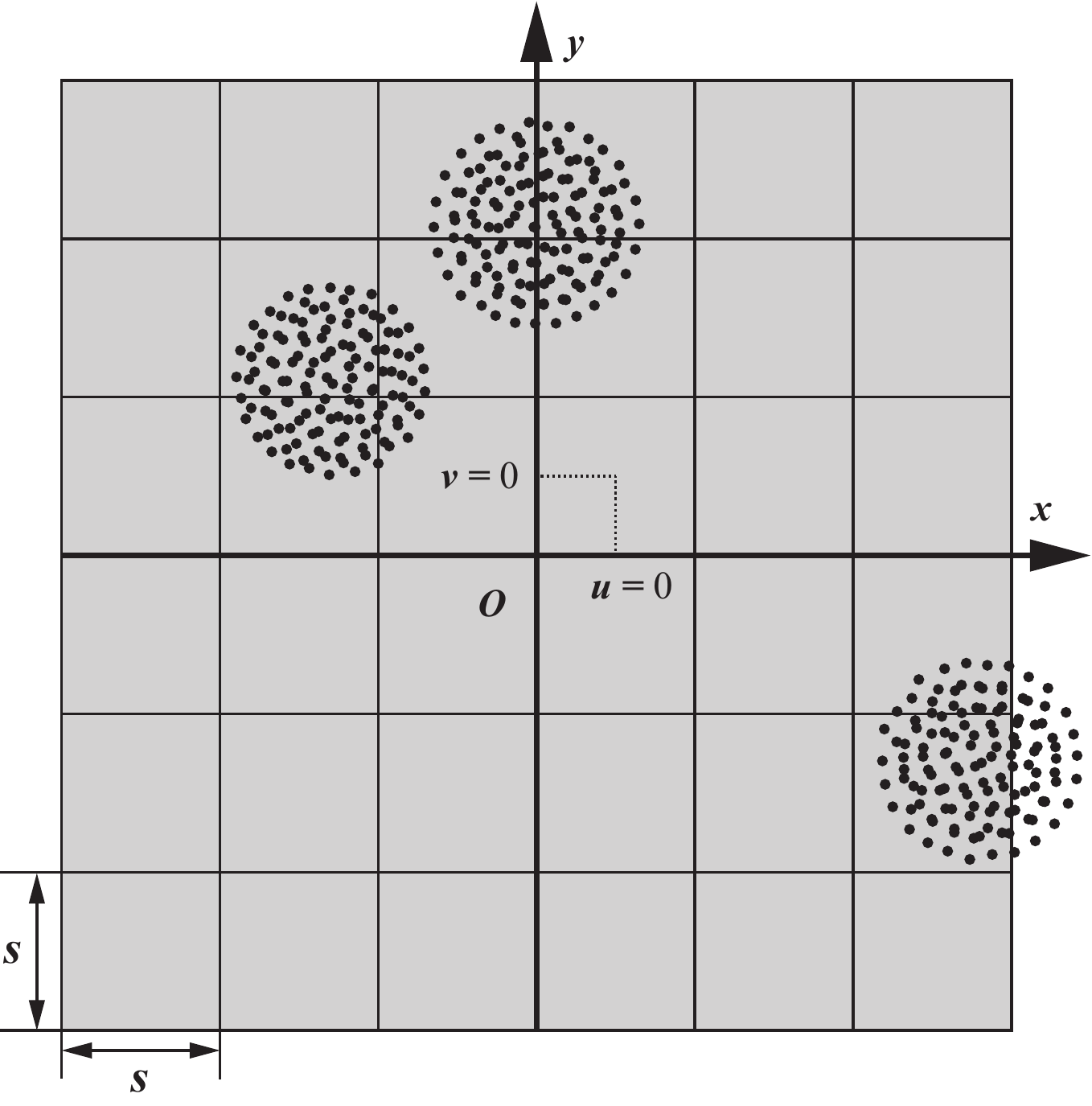}
	\caption{Image rendering. The coordinates of the pixels are counted from $(u,v)=(0,0)$.}
	\label{fig:figure04}
\end{figure}

The image rendering process evaluates the value of each pixel using the information stored in the photon map and the parameters of the detector, which includes image plane sampling, the conversion from photon to ADU, adding bias, dark and flat-field effect, manufacturing defective pixels. The process is illustrated by Figure~\ref{fig:figure04}. For a CCD with pixel size $s$ in the unit same as $\tilde{x}$ and $\tilde{y}$, the number of photons from sources falling into the pixel $(u,v)$ can be obtained by sampling the image coordinates of photons $(\tilde{x}_{ij}, \tilde{y}_{ij})$. There is
\begin{equation}
	E_{\rm source}(u,v) = \rm{card}\it \ \{(\tilde{x}_{ij}, \tilde{y}_{ij})|\ \langle \frac{\tilde{x}_{ij}}{s} \rangle=u, \langle \frac{\tilde{y}_{ij}}{s}\rangle=v\},
\end{equation}
where $\langle\ast\rangle$ means the integer part of a number $\ast$, and the operator $\rm{card} \{\ast\}$ means the number of elements of the set $\{\ast\}$. As illustrated by Figure~\ref{fig:figure04},  photons from the same source may fall on different pixels. A single-pixel may receive photons from different sources. A part of photons from a source may fall outside the CCD plane. The discrete shape of the CCD pixels determines that it can record the photon's approximate position.

Strictly speaking, the photons coming from the sky background go through the same imaging process as those from sources. But for a telescope with a small field of view, the sky background can be considered approximately uniform. It can be rendered by adding a constant value $E_{\rm back}(u,v)$ in photon-electrons to all pixel value. Each pixel value $E_{\rm back}(u,v)$ should be replaced with a random number of events drawn from a Poisson or Gaussian distribution with mean $E_{\rm back}(u,v)$ to mimic the noise.  Alternatively, the background $E_{\rm back}(u,v)$ can be estimated from observed image. The sum number of photon-electron 
\begin{equation}
	E(u,v)=E_{\rm source}(u,v)+E_{\rm back}(u,v)
\end{equation}
It gives a pure photon-electron image not contaminated by the readout process, which is very useful in image processing. Because the various calibrations (gain, bias, dark, flat, charge transfer inefficiency) do not have to be performed. However, if we want to obtain an image containing various effects, just like the real CCD readout image, the whole digitization process can be realized by
\begin{equation}
	ADU(u,v) = \frac{E(u,v)}{Gain}\cdot Flat(u,v)+Bias(u,v)+Dark(u,v),
\end{equation}
where $Gain$, $Flat(u,v)$, $Bias(u,v)$, and $Dark(u,v)$ are the gain of CCD, flat, bias, and dark, respectively. Their values can be arbitrarily specified or measured using an actual device.

Some artificial defects can be added to make the images more realistic, such as hot pixels and dead pixels.  Hot pixels can be added to the image by randomly choosing a fraction of the pixels and placing electrons equal to the full well depth. Similarly, a fraction of the pixels can be flagged as dead and remove those pixels' electrons. The same hot pixels and dead pixels should be added to every image for a particular version of the simulation using the same random number seed for all observations. 

\subsection{Image attribution}
From the photon map and the ADU-based image, we can easily deduce some useful parameters of each source in photon-electron or ADU, such as flux, barycenter, and signal-to-noise ratio. These parameters are essential as standard answers when using the simulated images as test data set for the image processing algorithm.

\section{Results}\label{sec:results}
To test the presented method, we compared the simulated images with the observed ones. The observed images were obtained with the one-meter telescope at Weihai Observatory of Shandong University. The telescope has an f/8 classic Cassegrain design with a field of view of $12^{\prime}\times12^{\prime}$. The photometric system, the CCD camera, and the site astro-climate are introduced in detail by~\citeasnoun{ShaoMingHu.2014}. The simulation architecture was implemented based on some off-the-shelf libraries, such as Astropy~\cite{Astropy.2013,Astropy.2018}, Skyfield~\cite{Rhodes.2019}, and Astroquery~\cite{Ginsburg.2019}. A demo script written in Python code is available on the author's Github~\footnote{\url{https://github.com/dujunju}}. The similarities between the simulated images and the observed ones are measured by residual and Pearson correlation coefficient.  

\begin{figure}[ht!]
	\centering
	\includegraphics[width=1.0\columnwidth]{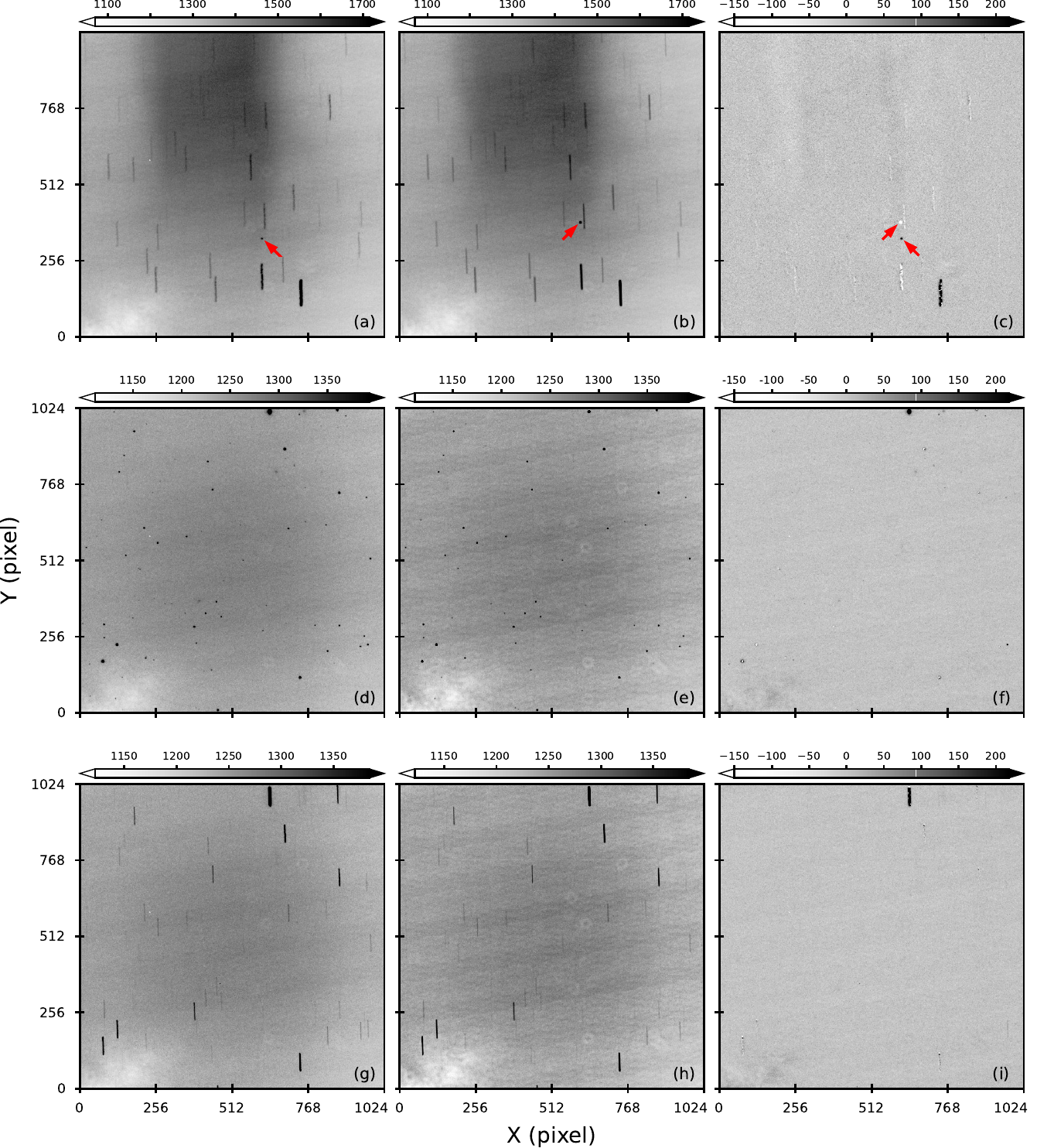}
	\caption{Images obtained in three tracking modes. The left column shows the observed images obtained in (a) target tracking mode, (d) sidereal tracking mode, and (g) parking mode. The middle column ((b)(e)(h)) presents the corresponding simulated images. The right column ((c)(f)(i)) illustrates the residuals between them. Some bright halo in the middle part of (a) may because the shutter is not entirely closed, and there is some light leakage during the readout process.\label{fig:figure05}}
\end{figure}

Tracking mode is the first issue to be identified. Three observed images obtained in different tracking modes, the corresponding simulated images, and the residuals between them are shown in Figure~\ref{fig:figure05}. Figure~\ref{fig:figure05} (a) is a raw observed image of a GEO space debris in target tracking mode. An arrow indicates a point-like target image. Figure~\ref{fig:figure05} (d) and (g) are raw observed images of BL Lac 3C 66A (02$^{\rm h}$22$^{\rm m}$39.61$^{\rm s}$ ~$+43^{\circ}02^{\prime}07.80^{\prime\prime}$) in sidereal tracking mode and parking mode, respectively. Limited by the small field of view and our telescope's tracking mechanism, it is not easy to obtain relevant observations of moving targets in sidereal or parking mode. Therefore, images of star fields were used instead to illustrate the latter two tracking modes. 

Figure~\ref{fig:figure05} (b)(e)(h) are the simulated images corresponding to (a)(d)(g). The simulation parameters were well tuning or estimated to restored the real observation scenes as far as possible. The parameters of filed stars are from UCAC4~\cite{Zacharias.2012} and Gaia EDR3~\cite{Gaia.2020}, while the parameters of the moving target are from Space-Track.org and our historical observations. The observation settings are readily available. Atmospheric and equipment parameters, including seeing, zero-point and extinction coefficient (or the difference between the instrument magnitudes and the standard magnitudes), plate constants, calibration images, and sky background, were estimated from actual observations.

The differences between the simulated images and the observed ones are indistinguishable by naked eyes. The residuals between them are shown in Figure~\ref{fig:figure05} (c)(f)(i), which are almost white noise. Only the images of individual stars show some residual traces, which may be due to the magnitude error in the catalog, the simplified PSF function, or the imperfect plate constants. As shown in Figure~\ref{fig:figure05} (b), the simulated target image is not aligned with the observed one, nor is it in the center of the frame. The former is due to the orbital prediction error, while the latter is due to the telescope pointing deviation. The correlation coefficients between the simulated and observed images in target, sidereal, and parking tracking mode were 0.9512, 0.9648, and 0.8880, respectively. The correlation coefficients of the calibrated images were also calculated to exclude the bias and flat's possible effects—the correlation coefficients became 0.9125, 0.9652, and 0.8968, respectively.

The results presented in Figure~\ref{fig:figure05} are just an random example. To prove this point, we selected a continuous period of observation images to analyze. As mentioned above, our telescope is mainly suitable for follow-up observation in target tracking mode. Therefore, we only compare a large number of simulated images with observed ones in target tracking mode.   

\begin{figure}[ht!]
    \centering
	\includegraphics[width=0.6\columnwidth]{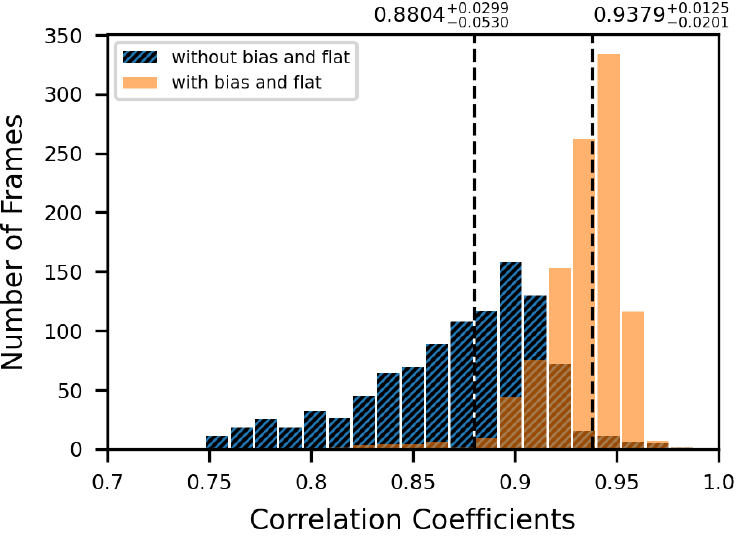}
	\caption{Histogram of the correlation coefficients between 1020 observed images and corresponding simulated ones.}
	\label{fig:figure06}
\end{figure}

We observed a GEO space debris in target tracking mode on 2021 January 29 and obtained thousands of photometric images in the V band. But most of them have few field stars. We did sources extraction using SExtrator~\cite{Bertin.1996}, and selected 1020 images containing more than five field stars. In response, we generated 1020 simulated images.  Correlation coefficients between observed images and corresponding simulated ones are calculated. The histogram of correlation coefficients is shown in Figure~\ref{fig:figure06}. The median correlation coefficient is $0.9379_{-0.0201}^{+0.0125}$.  After calibration, the median correlation coefficient became $0.8804_{-0.0530}^{+0.0299}$.  These values indicate that our method can generate photometric images with very high fidelity.

\begin{figure}[ht!]
	\centering
	\includegraphics[width=1.0\columnwidth]{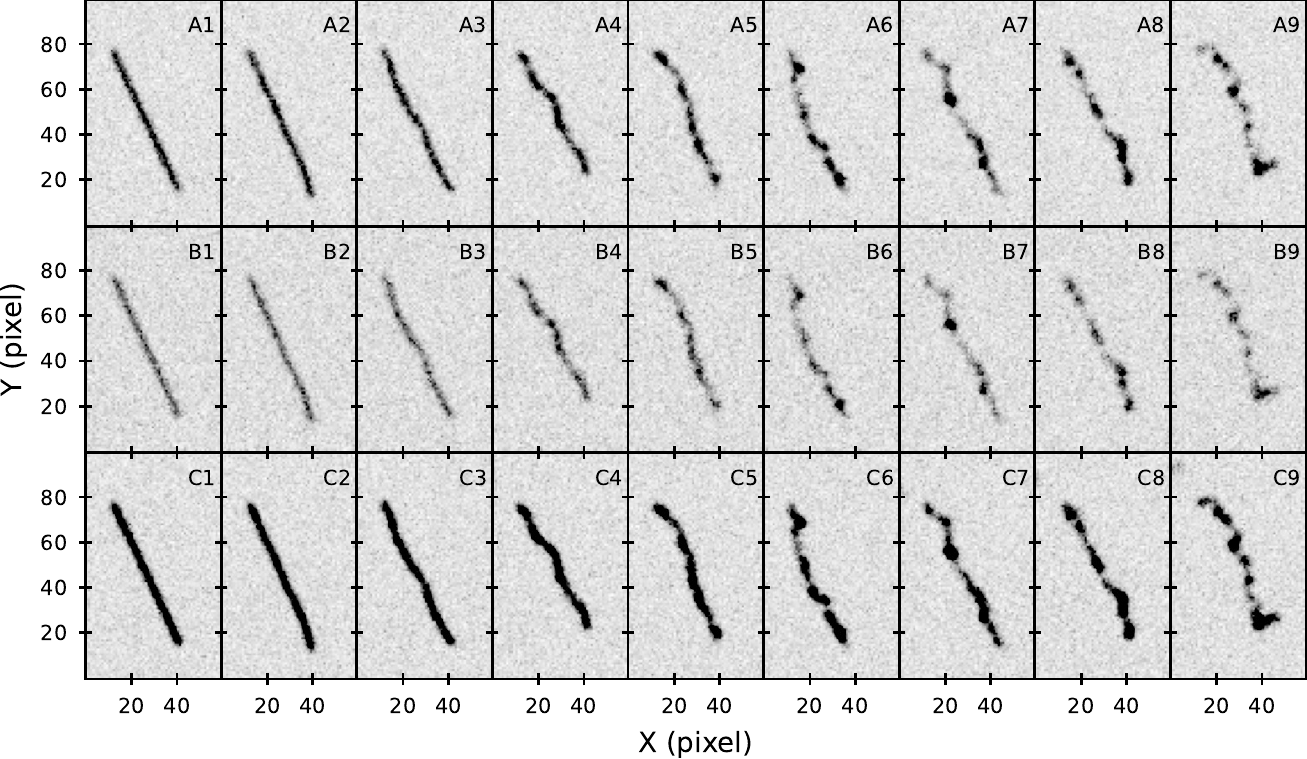}
	\caption{Screenshots of three streak-like images (A, B, C) in simulated images ($1,2,3,4,5,6,7,8,9$) generated with different distortion $\sigma_{\rm A}=\sigma_{\rm D} = 0.0, 1.0, 2.0, 3.0, 4.0, 5.0, 6.0, 7.0, 8.0$ in  $10^{-4}$ deg $\cdot$ s$^{-0.5}$  from left to right.\label{fig:figure07}}	
\end{figure}

The ability to generate distorted source images and provide them for the testing of image processing algorithms is another powerful feature of our method. Distortions of source images are common in practical observations. In our method, the distortion of the source images can be easily adjusted by $\sigma_{A}$ and $\sigma_{D}$.  For example, images with different degrees of distortion are shown in Figure~\ref{fig:figure07}. The shape of streak-like images changes unpredictably from frame to frame and cannot be expressed in an analytic form accurately. But all streak-like images within a single frame share similar shapes, which means the telescope instability affects all field stars in the same way. This phenomenon is consistent with the actual observation shown in Figure~\ref{fig:figure02}, which proved that the distortion of the source image was due to the instability of the telescope.

\section{Conclusions and discussion}\label{sec:discussion}
In this paper, we provide a new strategy to simulate the photometric images of moving targets. Although many authors have made significant contributions and achievements in this field, such as SkyMaker~\cite{Bertin.2009} and PhoSim~\cite{Peterson.2015}, our approach still shows extraordinary strength. First, this method's whole implementation process is tracing each photon's journey, which is similar to the actual imaging process, so each step is easy to understand. Second, the propagation path of each photon is determined by the statistical law under various factors, including the distribution of photon arrival time, the random jitter of the telescope, and the point spread function. Compared to the convolution-based method~\cite{Kouprianov.2008,Veres.2012}, our method can better represent the discrete distribution of photons. Third, this method is easy to implement and use, and it can easily generate images under different observation scenarios by adjusting parameters. Our method demonstrates high efficiency, accuracy, and flexibility.

An observation scene consisting of hundreds of sources might involve hundreds of millions of photons. Determining the propagation path of a vast amount of photons requires a large amount of computation, restricting the simulation speed. Experiments show that it takes about a minute to produce an image containing 20 million photons. Using parallel computing, optimizing code, or using more efficient programming languages, the speed can be greatly increased. On the other hand, some subtle factors that may influence the simulation have not been considered, such as the PSF changes in different regions of the image, readout noise, blooming of source, and the structure of extended sources. These factors can be added to the framework in the future depending on the research needs.  

Image simulation is a useful tool to study the photometry of moving targets, and it will be continually developed. We believe that our method will play a particular role in making observation plans, interpreting observation images, and providing test images to develop and test image processing algorithms.

\ack
This work is supported by the Natural Science Foundation of China under Grant No. 11873035, the Natural Science Foundation of Shandong province (No. JQ201702), and the Young Scholars Program of Shandong University (No. 20820162003). Thanks, Sofya Alexeeva, for revising this paper. 

\small
\bibliographystyle{dcu}
\bibliography{references}

\end{document}